\def\cm{\ifmmode \,\, {\rm cm} \else sec \fi}
\def\sr{\ifmmode \,\, {\rm sr} \else sec \fi}
\def\s{\ifmmode \,\, {\rm s} \else sec \fi}
\def\sec{\ifmmode \,\, {\rm sec} \else sec \fi}
\def\eV {\ifmmode \,\, {\rm eV} \else eV \fi}
\def\keV{\ifmmode \,\, {\rm keV} \else keV \fi}
\def\MeV{\ifmmode \,\, {\rm MeV} \else MeV \fi}
\def\GeV{\ifmmode \,\, {\rm GeV} \else GeV \fi}
\def\TeV{\ifmmode \,\, {\rm TeV} \else TeV \fi}
\def\fm{\ifmmode \,\, {\rm fm} \else TeV \fi}
\def\pbarn{\ifmmode \,\, {\rm pb} \else pb \fi}
\def\deg{\ifmmode ^\circ\, \else $^\circ\,$ \fi}
\def\km{\ifmmode {\rm km}\, \else km \fi}
\def\Mpc{\ifmmode {\rm Mpc}\, \else Mpc \fi}
\def\Gyr{\ifmmode {\rm Gyr}\, \else Gyr \fi}

\def\sr{\ifmmode \,\, {\rm sr} \else sr \fi}

\def\picture #1 by #2 (#3){
  \vbox to #2{
    \hrule width #1 height 0pt depth 0pt
    \vfill
    \special{picture #3} 
    }
  }
\def\scaledpicture #1 by #2 (#3 scaled #4){{
  \dimen0=#1 \dimen1=#2
  \divide\dimen0 by 1000 \multiply\dimen0 by #4
  \divide\dimen1 by 1000 \multiply\dimen1 by #4
  \picture \dimen0 by \dimen1 (#3 scaled #4)}
  }
\def\boxit#1{$$\vbox{\hrule\hbox{\vrule\kern3pt
     \vbox{\kern3pt \hbox to 13cm{\hfil }\vskip
     #1cm\kern3pt} \kern3pt \vrule}
     \hrule}$$}

\documentclass{ws-p9-75x6-50}

\begin{document}
 \hspace{10cm}  {\sl Rom2F/2001/06}
\vskip 1truecm
\title{ Exotic Matter research in Space}

\author{P.Picozza, A. Morselli}

\address{Dept. of Physics,  Univ. of Roma ''Tor Vergata'' and INFN, Sezione di Roma 2, Roma,
Via della Ricerca Scientifica 1, 00133,  Italy \\E-mail: picozza@roma2.infn.it, morselli@roma2.infn.it}


\maketitle

\abstracts{
The direct detection of annihilation products in cosmic rays offers  an
alternative way to search for dark matter particles candidates. Here we will
see in particular that the study of the spectrum of antiproton and positrons
  offers good possibilities to perform this search and we will
review our experimental effort in this direction.
}

\section {Probing for  dark matter with cosmic rays space experiments}

 Galactic dark matter was suggested to solve the discrepancy between observed
(luminous) matter in the Universe and that inferred from dynamical considerations.
A  matter density of about
30 - 40\% of the critical density of the Universe can be composed of dark matter. One possible form of
dark matter could be weakly interacting massive particles (WIMPs) and a good candidate for WIMP's 
is the Lightest Supersymmetric Particle (LSP) in R-parity conserving SUSY models. In  most of the  Supersymmetric
theories the LSP is the neutralino $\chi$ that is a combination of the partners of the $\gamma$, $Z$ and the
neutral Higgs particles (see \cite{glast_gros} and references threin). 
 Neutralinos are Majorana fermions and will annihilate with each other
in the halo producing leptons, quarks, gluons, gauge bosons and Higgs
bosons. The quarks, gauge bosons and Higgs bosons will decay and/or
form jets that will give rise to antiprotons (and antineutrons which
decay shortly to antiprotons). 

The idea of exploiting cosmic antiprotons measurements to probe
unconventional particle physics and astrophysics scenarios has a long history \cite{ras} and 
was stimulated by early reports \cite{golden} of unexpectedly large values for the 
antiproton to proton ratio in cosmic rays.

 Quantitative estimation of the antiproton flux due to dark-matter annihilation requires
assumptions about the WIMP mass and dark matter density. 

The  solar modulation   introduce incertanties in the
antiproton spectrum at low energies (below 1 GeV), so the best place to look from a deviation 
of the spectra is at high energies where there can be a 
neutralino signal  rather sharply peaked at an energy higher than the 
maximum in the background  if its spectrum   decreases rapidly 
at low energies. This effect can be produced 
by high mass neutralinos with negligible branching ratio into 
$b \bar{b}$ or $t \bar{t}$, which is the case e.g. for a very pure 
heavy Higgsino-like neutralino. 

\begin{figure}
  \begin{center}
    \mbox{\epsfig{file=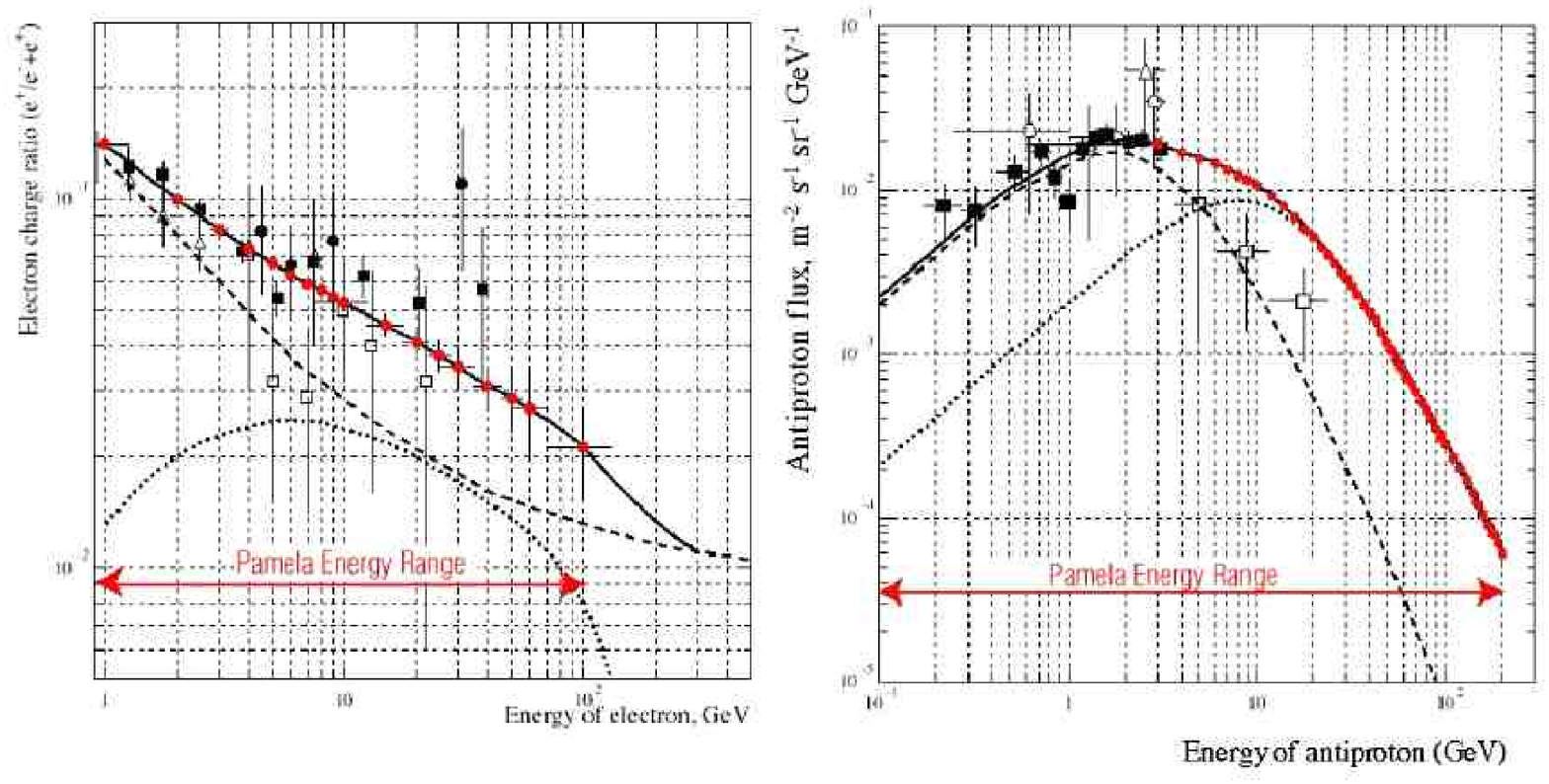,height=8cm,width=12.6cm} }
  \end{center}
    \caption[]{Distortion of the secondary 
positron fraction (on the left) and  secondary antiproton flux (on the right) induced by a signal from a
heavy neutralino}
    \label{antip}
\end{figure}

In figure \ref{antip} (on the right)
there are the  experimental data   for the antiproton flux  \cite{antip_data} together with the 
distortion on the antiproton flux (dashed line) due to one possible
contribution from neutralino annihilation (dotted line, from \cite{Ullio}). 
Total expected flux is shown
by solid line. The antiproton data that PAMELA 
would obtain in a single year of observation for
one of the Higgsino annihilation models are shown
by grey circles. The PAMELA experiment will be described in the next section.

 Exotic dark matter can also be investigated with
positrons. Electrons are a primary component of cosmic rays and emerge from ambient material in an acceleration
process that may or may not be identical to that which generates energetic nucleons. Positrons are thought to be
mostly secondary in origin. WIMPs could contribute to the positron flux by direct annihilation into e$^+$ e$^-$,
and to continuum positrons from the other annihilation channels \cite{pam_e}. This will be seen as a
excess or bump beginning at a few GeV and extending upward in energy to a point depending on the WIMP mass.
Current experimental data is not sufficient to make solid conclusions, but there is some evidence for a positron
excess \cite{el_excess}.

In figure \ref{antip} (on the left) there are the experimental data \cite{el_data} for  the  positron
fraction  together with the distortion of the secondary positron fraction (dashed line) due to
one possible contribution from neutralino annihilation (dotted line, from \cite{pam_e}).  
The expected data
from the PAMELA experiment in the annihilation scenario for one year of operation are shown by grey
circles.

This effort will be complementary to a similar search with high-energy
gamma-ray instruments such as GLAST (10- 300 GeV) and ground-based air Cherenkov telescopes (100-1000 GeV)  
looking of possible signature of the existence of the LSP as a bump in the spectrum  of the diffuse gamma ray
background around the neutralino mass due to neutralino annihilation in the halo \cite{glast_gros}

\section{the PAMELA apparatus}

 PAMELA  is an satellite-borne magnet spectrometer built by the WiZard-PAMELA collaboration
\cite{pamela}. It will be installed on-board of the RESURS-5 ARTIKA satellite to be launched in the 2003 for a
mission at least three years long. The satellite orbit is polar, sun-synchronous and 700 km high.

 The list of the people and  the Institution involved in the collaboration together with the on-line
status of the project is available at {\sl http://wizard.roma2.infn.it/}.
\par
\begin{figure}
  \begin{center}
    \mbox{\epsfig{file=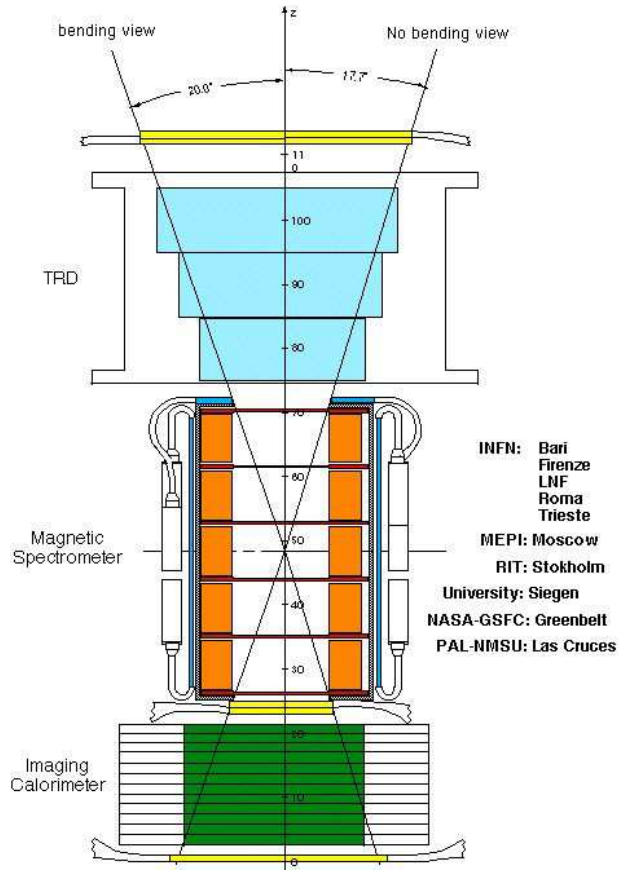,height=11.5cm} }
  \end{center}
    \caption[] {Schematic of the PAMELA  baseline instrument. }
    \label{pam}
\end{figure}

\noindent
The Pamela telescope, shown in figure \ref{pam},  consists of the following elements: 
a magnet + tracker system, an imaging calorimeter, a Transition-Radiation-Counter (TRD),
scintillation counter hodoscopes  for Time-of-Flight and Trigger, 
  an anticoincidence scintillation counter. 
The magnet + tracker system consist of 5 permanent magnets, each 8 cm high,
interleaving  6 detection planes of the silicon microstrip tracker. The whole closed in a
ferromagnetic screen and surrounded on its sides by a system of anticoincidence
   scintillation counters. The  resolution of the
tracking system is about 4 $\mu m$. The magnetic field
    inside the magnet will be $\sim$ 0.4 T, so the Maximum Detectable Rigidity (MDR)
will be around 800 GV/c.

 \subsection{ PAMELA Scientific Objectives}

The observational objectives of the PAMELA instrument are the measurement of the spectra of
antiprotons, positrons and nuclei in a wide range of energies, the search for
primordial antimatter and the study the cosmic ray fluxes over half a solar cycle.
Data gathered with the PAMELA
instrument will deal with a wide range of fundamental issues. These include: 
\vskip 0.2cm
\par\noindent   
$\bullet$ the role of Grand Unified Theories in Cosmology in relation to antimatter and dark matter. 
\par\noindent
$\bullet$ the understanding of the acceleration and propagation of cosmic rays. 
\par\noindent
$\bullet$ the role of solar, terrestrial and heliosperic relationships to energetic particle
propagation in the heliosphere. 
\vskip 0.2cm
The PAMELA observations will extend the results of balloon-borne experiments over an
unexplored range of energies with unprecedented statistics and will complement
information gathered from Great Space Observatories. These observational objectives can be
schematically listed in the following points:
\vskip 0.2cm
\par\noindent	
$\bullet$ Measurement of the energy antiproton spectrum in a large energy range: from 100 MeV up
to 150 GeV (present limits 0.4 - 20 GeV );
\par\noindent	
$\bullet$ Measurement of the energy positron spectrum in a large energy range: from 100 MeV up
to 200 GeV (present limits 0.7 - 30 GeV);
\par\noindent	
$\bullet$ Search for anti-nuclei with a sensitivity of 6 10$^{-8}$ in the anti-helium/helium ratio
(present limit about $10^{-5}$);
\par\noindent	
$\bullet$ Measurement of the electron energy spectrum up to 1000 GeV;
\par\noindent	
$\bullet$ Continuous monitoring of the cosmic rays solar modulation during and after the 23rd
maximum of solar activity;
	\par\noindent 
$\bullet$ Studies of the time and energy distributions of the energetic particles emitted in
solar flares.
\vskip 0.2cm

The low energy antiproton and positron measurements and the last two objectives are peculiar of the
PAMELA experiment because the satellite travels in a polar orbit. It spends a large fraction of its
time in the high latitude and Polar Regions, where the cut-off due to the terrestrial magnetic field
is negligible.

The scientific relevance of these objectives is enhanced by the length of the mission, that is
planned to last not less than three years, but could be prolonged for many other years because of
the orbit altitude and the maximization of the electric power due to its sun-synchronism.

In the following sub-sections the above objectives are discussed in some detail. 

$\bullet${ Antiprotons}

Antiprotons have been observed in the cosmic rays since 1979 by balloon-borne experiments; prior
to these measurements it was generally expected that all primary cosmic rays experienced the same
basic history during their acceleration and propagation. Similarly it was assumed that all secondary
were produced and stored in the same regions of the Galaxy. These assumptions were quite adequate to
explain the observations of secondary Z $\geq$ 3 nuclei in cosmic rays. 

Because of atmospheric backgrounds and limited flight times, balloon-borne experiments can only
measure the energy spectrum of antiprotons up to about 20 GeV. In addition, the sensitivity of
the balloon observations is limited by the difficulty in eliminating large fluxes of atmospheric
secondary particles. PAMELA will be able to measure the energy spectrum of antiprotons up to 150
GeV. 

$\bullet${ Positrons and Electrons.}

Positrons and electrons are unique among cosmic rays because they are the lightest charged leptons.
Due to their low mass, high-energy electrons and positrons undergo interactions with the ISM,
which result in severe energy losses at high energies. While most of the observed electrons are
believed to be of primary origin, the origin of positrons is yet to be established. Positrons are
even harder to observe than antiprotons due to the high flux of protons (more than 1000 times
higher). All observations to date have suffered from the risk of subtracting significant background.
The majority of data shows an excess of positrons above the flux expected by the simple leaky-box
model and may even indicate a rise in the positron/electron ratio at energies greater than 15 GeV.
These direct observations, combined with the observation of a positron annihilation emission line
from the galactic disk and the high antiproton fluxes, give rise to questions such as:
\vskip 0.2cm
\par\noindent
1.	Are there positrons in the cosmic rays that are not produced as secondary? 
\par\noindent
2.	Is there a relationship between the antiproton and positron excesses? 
\par\noindent
3.	If positrons are indeed all secondary particles, at what point do the radiative
losses become important?
\vskip 0.2cm \par\noindent
Observation of positron over a very large energy range should yield new insights into galactic
processes. In particular, as we showed, the signature of WIMP particle existence in Dark Matter could be found in
the high-energy spectrum of positron. As for antiprotons, PAMELA will aim to measure accurately the
spectra of positrons from low (cut-off) energies up to the highest energies attainable (about 200
GeV). Furthermore, together with the measurement of the spectra of electrons (up to about 1000 GeV)
PAMELA can provide information on the:
\vskip 0.2cm
\par\noindent
1.	Acceleration of electrons and the distribution of acceleration sites; 
\par\noindent
2.	Cosmic-ray lifetime, and the physical conditions in the containment volume; 
\par\noindent
3.	Magnitude of re-acceleration by interstellar shock waves.
\vskip 0.2cm

$\bullet${ Search for antimatter.}

Detection of antimatter of primary origin in cosmic rays would be a discovery of fundamental
significance. Cosmic-ray searches that have been made so far have yielded only upper limits of one
part in 10$^{-4}$ for heavy nuclei (Z$>$2) and one part in $10^{-5}$ for helium \cite{anti_he}. The detection
of anti-nuclei in cosmic rays would provide direct evidence of the existence of antimatter in the
universe. Baryons and photons were produced in the Big Bang in equal amount, but from observation of
the 2.7 K cosmic background radiation and the present matter density of the universe we know that
only about one baryon remains for ever 10$^9$ photons. The current theory suggests that the remaining
matter is the remnant of the almost complete annihilation of matter and antimatter at some early
epoch, which stopped only when there was no more antimatter to annihilate. Starting from a
matter/antimatter symmetric Universe, the required conditions for a following asymmetric evolution
are the CP violation, the baryon number non-conservation and a non equilibrium environment. On the
basis of gamma-ray observations, the coexistence of condensed matter and antimatter on scales
smaller than that of clusters of galaxies has been virtually ruled out. However, no observations
presently exclude the possibility that the domain size for establishing the sign of CP violation is
as large as a cluster or super-cluster of galaxies. For example, there could be equality in the
number of super-clusters and anti-super-clusters. Similarly, there is nothing that excludes the
possibility that a small fraction of the cosmic rays observed at Earth reach our Galaxy from nearby
super-clusters. PAMELA will search for anti-nuclei with sensitivity on the anti-helium/helium ratio
of some units in
$10^{-8}$.

$\bullet${Additional objectives}

The continuous determination of the direction, latitude and longitude of the primary electrons,
positrons, protons, antiprotons and light nuclei during several years and over a large energy range
(100 MeV up to several tens GeV) will provide a great opportunity to investigate other scientific
issues. The PAMELA experiment will be able to address these additional objectives besides the
primary ones above described. Indeed, during its orbiting around the Earth, the satellite will
encounter all that kind of events that are related to the solar activity and to the terrestrial
geomagnetic effects. 

The analysis of all data gathered over its mission will provide a significant complement to the
measurements performed so far with dedicated experiments of different concepts. The additional
objectives of PAMELA are the following:
\vskip 0.2cm
\par\noindent 1.	Modulation of galactic cosmic rays in the heliosphere;
\par\noindent 2.	Solar flare particle spectra;
\par\noindent 3.	Distribution and acceleration of solar cosmic rays (SCR's) in the internal
heliosphere;
\par\noindent 4.	Magnetosphere and magnetic field of the Earth;
\par\noindent 5.	Stationary and disturbed fluxes of high energy particles in the Earth's
magnetosphere;
\par\noindent 6.	Anomalous component of cosmic rays.
\vskip 0.2cm \par\noindent
	The PAMELA mission will carry out its observations during and after the maximum of the 23rd solar
activity cycle. The modulation effect on electrons, positrons, protons and nuclei during this period
will be investigated in order to find any dependence on charge sign, energy (rigidity) and mass of
the particles. Latitude and longitude distributions of the observed fluxes will be analysed as a
function of the solar activity to look for possible correlation. 

Most of the satellite measurements on the composition and spectrum of solar energetic particles from
the flares are limited to energies below a few tens of MeV. Experiments carried by GOES-7, SAMPEX
and  NINA 1 \&2 telescopes\cite{nina}, could determine the spectrum of heavy nuclei to about 100
MeV/n for a few flares. Using PAMELA it is possible to measure the energy spectrum of nuclei from
helium to at least oxygen from about 100 MeV/n up to an energy where the solar energetic particles
can be distinguished from the galactic cosmic rays. Solar flares with generation of similar
particles are rare but, as of 2002, when the solar activity is at its maximum, about ten of such
events per year will be available. The composition and the spectral shape of nuclear components
could vary from flare to flare and it would be very exciting to relate these variations to the
physical conditions in the flare sites.

For the first time, there will be an opportunity to observe at the same time high energy solar
particles with different charges and different masses. The observation of such particles will not
only allow to choose between different SCR's acceleration processes on the Sun, but also to gather
information on the distribution and acceleration processes of the particles in the internal part of
the heliosphere. The observation of SCR's, the determination of energy and temporal distributions of
various components will allow to carry out the internal tomography of the heliosphere: the shock
wave propagation and the fluxes of solar wind drivers (coronal mass ejection) will be investigated.

The experimental data obtained by PAMELA both in conditions of quiet and disturbed Earth's
magnetosphere will allow to study the variations of current systems and their influence on the
trajectory of cosmic rays and on the geomagnetic rigidity thresholds.

The altitude of the PAMELA orbit is about 700 km. For almost one third of its orbits, the instrument
will cross the internal part of the Earth radiation belt (the so called Brazilian anomaly region)
giving the opportunity of observing both primary cosmic rays and the high energy particles of the
Earth radiation belt itself.

The anomalous component of cosmic rays consists of partially (or single) ionised interstellar
neutral atoms accelerated at the termination shock and penetrating inside the heliosphere. Their
energy can vary from 10 MeV up to several hundred MeV which is sufficient to reach the vicinity of
the Earth and to be observed by satellites at high latitudes or outside the magnetosphere. Due to
the inclination of the PAMELA's satellite orbit (98 degrees) there is an opportunity to carry out
measurements of the anomalous component of cosmic rays both at high latitudes and in the radiation
belt of the Earth.

\subsection{ The PAMELA telescope.}

The PAMELA telescope design is well defined and completed. In the past two years prototypes of
detectors and sub-systems of the telescope have been built and tested, measuring their performances.
Therefore, the expected behaviour of the whole PAMELA experiment is well known and the required
techniques already defined and tested.

The concept of the PAMELA telescope is the same on which the proposed WIZARD experiment was based:
\vskip 0.2cm
\par\noindent 1.	A magnetic spectrometer to determine the sign of the electric charge with a very
high confidence degree, and to measure the momentum of the particles up to the highest energies for
which a useful flux of rare particles (as antiprotons and positrons) can be collected; 

\par\noindent 2.	An imaging calorimeter that can give, besides the measurement of the energy
released by the interacting particle (and indeed of the extra energy released in an annihilation
event), the pattern of the interaction of the particle inside the calorimeter, in order to identify
the particle itself. This last point, observing the annihilation pattern of antiprotons and possible
anti-nuclei, allows confirmation of their identity on an "event by event" basis; 

\par\noindent 3.	A velocity measurement system to
help the calorimeter in the identification of the nature of the particle;

The technical choices had to be made taking into account the mass and power limits of the payload,
dictated by the launch opportunity. They are the following:

	Permanent magnet system and micro-strip silicon sensors for the magnetic spectrometer. The magnetic
system is composed by five permanent magnet of Nd-Fe-B, each 8 cm high, that provide a field inside
the tracking volume of about 0.4 T. There are six planes of silicon micro-strip detectors and the
tests made in the past years on PAMELA detector prototypes resulted in a spatial resolution of 2.9
$\mu$m. Therefore, by considering a conservative approach, the spatial resolution requirement of the
PAMELA tracker is fixed to 4 $\mu$m;

	Silicon strip sensors for the imaging calorimeter, interleaved with tungsten plates as absorber;
this choice minimizes the volume of the calorimeter, and maximizes indeed its geometrical
acceptance; the high granularity assumed in PAMELA allows a very good separation between
electromagnetic showers and interacting or not interacting hadrons; the chosen depth of 16 radiation
length allows a good resolution in the measurement of the energy of electromagnetic particles,
further extending the energy spectrum measurements of electrons and positrons;

	A Transition Radiation Detector (TRD) for distinguishing electromagnetic particles from hadrons up
to very high energy (about 1000 GeV); the TRD is based on small diameter straw tubes arranged in
double layer planes interleaved by carbon fibre radiator. The use of the straw tubes allows to
perform many energy loss measurements along the particle trajectory bringing the selection
capability of the instrument down to less than 1 GeV, and to track all particles before their
entrance in the magnetic spectrometer, cleaning the sample of the particles accepted at its entrance;

	Several (6) scintillation counter hodoscopes (each 7 mm thick) for the construction of the triggers
and for the TOF measurements; the use of several hodoscopes allows independent TOF measurements,
improving the precision and the safety; 

	Finally a set of scintillation counters covering the top edge and the sides of the magnetic
spectrometer and the bottom part of the calorimeter completes the telescope, for a further labelling
of contaminating events.

The main characteristics of the PAMELA experiment are:
\vskip 0.2cm
\par\noindent 1.	A Maximum Detectable Rigidity (MDR) of 800 GV/c;

\par\noindent 2.	An acceptance (geometrical factor) of 20.5 cm$^2$ sr; 

\par\noindent 3.	A maximum energy reachable in the antiproton spectrum measurement of 150 GeV or
greater, depending on the content of antiprotons in cosmic rays; 

\par\noindent 4.	A maximum energy reachable in the electron and positron spectrum measurement respectively of
1000 and 200 GeV;

\par\noindent 5.	A sensitivity for the anti-helium
search of 6  $10^{-8}$ in the anti-helium/helium ratio; 

\par\noindent 6.	A total volume of $90{\rm x} 90{\rm x}110$ cm$^3$;

\par\noindent 7.	A total mass of 395 kg;

\par\noindent 8.	A number of read-out channels of 43498;

\par\noindent 9.	A power consumption of 267 W.

\subsection{ The PAMELA rates and data flow.}

The PAMELA experiment will be put in polar orbit; therefore the total number of collected events
will be high in spite of the small Geometric Factor (GF): the magnetic cut-off due to the Earth
magnetic field nearly cancels on the poles, while at medium and low latitudes the bulk of cosmic
rays is cut out by the Earth magnetic field, that acts for them like a mirror.

The total rate of particle collected by the PAMELA telescope, averaged on the polar orbit and on the
solar activity cycle, is 3.3 event/s, i.e. about 3  10$^5$ event/day (mainly protons),
ranging from 2.3 event/s to 4.2 event/s going from the period of maximum solar activity to its
minimum. Averaging on the solar activity of the first three years foreseen for the PAMELA flight
(2002-2005) we expect to collect in 10$^8$ s the following approximate numbers for particles,
antiparticles and some nuclei:

\begin{table} [h]         
\begin{center}
\begin{tabular}{|l l|l l|}\hline 
	protons	   &  3 $ 10^8 $   & anti-protons &  3 $ 10^4 $ \\ \hline
	electrons	 &  3 $ 10^6 $   & positrons    & 	1 $ 10^5 $ \\ \hline
	He nuclei	 & 	4 $ 10^7 $   & Be nuclei    & 	4 $ 10^4 $ \\ \hline
	C nuclei	  &  4 $ 10^5 $   & anti-nuclei limit &  6 $ 10^{-8}$  (90\% C.L.)\\ \hline
\end{tabular}
\end{center}
\end{table}

The expected information flow from PAMELA is of the order of 2.5 kB/trigger; with not more than
5 $10^5 $ trigger/day, this results in a maximum amount of information of 1.25 GB/day.

\subsection{ The mission profile and the spacecraft RESURS-5 ARKTIKA.}

The PAMELA experiment will be installed on the up-ward side of the RESURS-5 ARKTIKA satellite, that
will be continuously oriented down-ward to the Earth during all its mission, in order to fulfil a
program of Earth surface observation. Furthermore the satellite will travel in a quasi-circular,
about 700-km high, polar orbit. This is an optimal situation for the observation of cosmic rays:

 \par\noindent 1.	the up-ward orientation of the PAMELA telescope on board of the satellite is the
required direction to observe cosmic rays without interference with the Earth and keeping far away
from the telescope acceptance the showers produced by quasi horizontal cosmic rays on the terrestrial
atmosphere; these showers are responsible of the strong increase of the background at low zenith
angles;

\par\noindent 2.	as above mentioned, the polar orbit maximizes the cosmic ray collection rate and
also minimises the geomagnetic cut-off in a significant portion of the satellite trajectory
(important for all scientific issues related to the solar activity and to the terrestrial
geomagnetic effects);

\par\noindent 3. finally, the large height of the orbit insures a long permanence of the satellite
in space due to the very small effect of the atmosphere. The stabilisation of the satellite is
obtained by magnetic devices to make the best use of this situation without being dependent from the
possible shortness of fuel.

\end{document}